\documentclass[12pt]{article}
\usepackage{comment}
\usepackage{scicite}
\usepackage{graphicx}
\usepackage{times}
\usepackage{comment}
\usepackage{multirow}

\usepackage{amsmath}
\usepackage{amssymb}
\usepackage[dvipsnames]{xcolor}
\newcommand{\beq}{\begin{equation}}
\newcommand{\eeq}{\end{equation}}

\newcommand{\ket}[1]{| #1 \rangle}

\newcommand{\threejm}[6]{ \left(\begin{array}{ccc} #1 & #3 & #5\\
                                              #2 & #4 & #6
                                \end{array}
                          \right)}

\newenvironment{sciabstract}{%
\begin{quote} \bf}
{\end{quote}}

\title{Ergodicity breaking in rapidly rotating C$_{60}$ fullerenes} %\title{Onset of high order icosahedral tensor interactions in C$_{60}$ fullerenes} 
\author
{Lee R. Liu$^{1,2\ast}$, Dina Rosenberg$^{1,2}$, P. Bryan Changala$^{1,2\dagger}$, Philip J.D. Crowley$^{3}$,\\ David J. Nesbitt$^{1,2,4}$, Norman Y. Yao$^{3}$, Timur Tscherbul$^{5}$, Jun Ye$^{1,2\ast}$\\
\normalsize{$^{1}$JILA, National Institute of Standards and Technology }\\ 
\normalsize{and University of Colorado, Boulder, CO 80309}\\
\normalsize{$^{2}$Department of Physics, University of Colorado, Boulder, CO 80309}\\
\normalsize{$^{3}$Department of Physics, Harvard University, Cambridge, MA, 02135}\\
\normalsize{$^{4}$Department of Chemistry, University of Colorado, Boulder, CO 80309}\\
\normalsize{$^{5}$Department of Physics, University of Nevada, Reno, NV, 89557}\\
\normalsize{$^\dagger$Present Address: Center for Astrophysics,}\\
\normalsize{Harvard \& Smithsonian, Cambridge, MA 02138}\\
\normalsize{$^\ast$Corresponding authors; e-mail: lee.richard.liu@gmail.com, ye@jila.colorado.edu}
}
\date{}
\begin{document} 

% Double-space the manuscript.

\baselineskip24pt

% Make the title.

\maketitle 

% Place your abstract within the special {sciabstract} environment.

\begin{sciabstract}
Ergodicity, the central tenet of statistical mechanics, requires that an isolated system will explore all of its available phase space permitted by energetic and symmetry constraints. %s with many degrees of freedom tend to thermalize. 
Mechanisms for violating ergodicity are of great interest for probing non-equilibrium matter and for protecting quantum coherence in complex systems. For decades, polyatomic molecules have served as an intriguing and challenging platform for probing ergodicity breaking in vibrational energy transport, particularly in the context of controlling chemical reactions. %In particular, highly excited molecular vibrations tend to efficiently redistribute their energy among many vibrational modes, while small amounts of vibrational energy tend to localize to a subset of vibrational modes~\cite{Sture1974,Leitner2018}. 
Here, we report the observation of \textit{rotational} ergodicity breaking in an unprecedentedly large and symmetric molecule, $^{12}$C$_{60}$. This is facilitated by the first ever observation of icosahedral ro-vibrational fine structure in any physical system, first predicted for $^{12}$C$_{60}$ in 1986~\cite{Harter1986}. The ergodicity breaking exhibits several surprising features: first, there are multiple transitions between ergodic and non-ergodic regimes as the total angular momentum is increased, and second, they occur well below the traditional vibrational ergodicity threshold~\cite{Logan1990}. These peculiar dynamics result from the molecules' unique combination of symmetry, size, and rigidity, highlighting the potential of fullerenes to uncover emergent phenomena in mesoscopic quantum systems.
\end{sciabstract}

% In setting up this template for *Science* papers, we've used both
% the \section* command and the \paragraph* command for topical
% divisions.  Which you use will of course depend on the type of paper
% you're writing.  Review Articles tend to have displayed headings, for
% which \section* is more appropriate; Research Articles, when they have
% formal topical divisions at all, tend to signal them with bold text
% that runs into the paragraph, for which \paragraph* is the right
% choice.  Either way, use the asterisk (*) modifier, as shown, to
% suppress numbering.

\paragraph*{One Sentence Summary:} High-sensitivity infrared spectroscopy of the 1185~cm$^{-1}$ spectral region in C$_{60}$ reveals peculiar rotational dynamics of a large symmetric molecule.
\section*{Main Text:} 
\textit{Introduction.} 
%Ergodicity is a central tenet of statistical mechanics, which states that systems with many degrees of freedom will efficiently redistribute energy amongst energetically accessible modes. %saying that a long time average can be replaced by a microcanonical ensemble average. It is a statement of the rate and extent of energy redistribution in the system, or the ability of the system to thermalize. Traditionally,  
%Understanding the mechanisms for breaking and restoring ergodicity is of great interest to explore the limits of classical thermodynamics, to control unimolecular reaction rates, or to generate robust non equilibrium states of matter, useful for protecting quantum coherence for quantum information processing or metrology. 
Isolated systems which break ergodicity have been explored in a variety of experimental settings, including spin glasses~\cite{Binder1986}, ultracold neutral atoms~\cite{Kinoshita2006,Schreiber2015} and ions~\cite{Smith2016}, and photonic crystals~\cite{Segev2013}. These systems exhibit ergodicity breaking of spin configurations and momentum or spatial distributions. By contrast, gas phase polyatomic molecules provide opportunities to probe the ergodicity breaking of collective (rotational and vibrational) excitations in a finite quantum system.  In this context, a topic of major interest has been the transport of energy deposited into molecular vibrations by optical pumping or collisions. Efficient intra-molecular vibrational redistribution (IVR) sets in once the vibrational energy exceeds a critical threshold defined by a combination of ro-vibrational coupling strengths and the density of states.~\cite{Nathanson1984, McCLELLAND1988,Sture1974,Logan1990,Leitner2018}. This vibrational ergodicity transition has been studied vigorously in the context of understanding and controlling unimolecular reaction dynamics~\cite{Nesbitt1996}. %IVR can be understood as in terms of an anderson localization transition in a 3N-6-1 dimensional disordered lattice with local tunnelling (vibrational coupling) matrix elements. 

Among polyatomic molecules, buckminsterfullerene ($^{12}$C$_{60}$) is notable for its structural rigidity and high degree of symmetry, which suppress IVR and allow for spectroscopic resolution~\cite{Changala2019} and optical pumping~\cite{Liu2022} of individual ro-vibrational states -- an unusual and fortuitous situation for a molecule with 174 vibrational modes. Its small rotational constant and stiff cage-like structure ensure that hundreds of rotational states are populated even when vibrational excitations are largely frozen out, achievable with modest buffer gas cooling to $\sim120~$K. Thus, %the large separation of rotational and vibrational energy scales ensures that 
a thermal ensemble of $^{12}$C$_{60}$ can reveal extensive, state-resolved rotational perturbations spanning hundreds of rotational quanta by eliminating vibrational ``hot bands". 

First observed and understood in atomic nuclei~\cite{Pavlichenkov1982,Pavlichenkov1988,Shimizu1989,Pavlichenkov1993,Frauendorf2001,Hubel2005,Kvasil2006,Kvasil2007}, rotational perturbations can arise from spherical symmetry breaking in the frame fixed to a rotating self-bound deformable body~\cite{Fox1977}, which lifts the degeneracy of different body-fixed projections of the total angular momentum vector \textbf{J}. %and belong to a general class of emergent behaviour characteristic of finite quantum systems~\cite{Pavlichenkov1993,Birman2013,Yannouleas2007}. 
Such perturbations, also called ``tensor interactions" due to their anisotropic nature, manifest in fine-structure splitting of the total angular momentum ($J$) multiplets in rovibrational spectra and encode rich dynamics such as rotational bifurcations~\cite{Pierre1989,Pavlichenkov1993}, as previously observed in tetrahedral SnH$_4$, CD$_4$, CF$_4$, SiH$_4$, and SiF$_4$ and octahedral
SF$_6$ molecules~\cite{Mcdowell1980,Patterson1982,Aldridge1975,Kim1979,Harter1984,Harter1978,Harter1979,Mcdowell1976,Borde1982}. %, notably in the context of %SnH$_4$, CD$_4$, CF$_4$, SiH$_4$, and SiF$_4$, similar patterns had been previously theorized and observed for other ,..., followed by gas-phase spherical top molecules such as tetrahedral CF$_4$ and SiF$_4$ and octahedral SF$_6$~\cite{Mcdowell1980,Patterson1982,Aldridge1975,Kim1979,Harter1984,Harter1978,Harter1979,Mcdowell1976,Borde1982}... Yet, 
Despite this, observing \textit{icosahedral} tensor interactions, first predicted for $^{12}$C$_{60}$ over three decades ago~\cite{Harter1986}, has remained an elusive goal, since there are far fewer examples of icosahedral molecules, non-spherical interactions occur only at higher orders of interactions, and icosahedral molecules are necessarily larger than lower-symmetry spherical top molecules, implying a smaller rotational constant. 

In this work, we observe these icosahedral tensor interaction splittings for the first time, revealing \textit{rotational} ergodicity transitions in $^{12}$C$_{60}$ at energies well below its IVR threshold~\cite{Logan1990}.  Specifically, as the molecule ``spins up" to higher \textit{J}, the dynamics of the angular momentum vector $\textbf{J}$ in the molecule fixed frame switches between ergodic and non-ergodic regimes. In the non-ergodic regime, distinct \textbf{J} trajectories exist in the same energy range, but remain separated by energy barriers. In the limit of high $J$, the tunnelling between these trajectories is too weak to restore ergodicity, leaving a characteristic signature in the fine-structure level statistics. This phenomenon differs from IVR in three key respects: (i) it involves the “transport” of the molecule frame orientation of \textbf{J} instead of vibrational energy, (ii) it can occur well below the IVR threshold, and (iii) it switches back and forth multiple times between ergodic (described by a 6th rank tensor interaction) and non-ergodic (described by a 10th rank tensor interaction) regimes as the angular momentum is varied.  As we will show, this peculiar dynamical behaviour arises from multiple avoided crossings with other vibrational states, which induce non-monotonic variations in the molecule's anisotropic character as $J$ is varied.

We describe the ro-vibrational structure of C$_{60}$ with a general field-free molecular Hamiltonian
\begin{equation} \label{eq:H_tot}
    H = H_\text{scalar}+ H_\text{tensor}
\end{equation}
The scalar Hamiltonian $H_\text{scalar}$ contains only those combinations of \textbf{J} and vibrational angular momentum $\boldsymbol\ell$ that preserve their spherical degeneracy~\cite{Hecht1961},
\begin{equation}\label{eq:H_scalar}
    H_\text{scalar} = \nu_0 + B\textbf{J}^2 + D\textbf{J}^4 + \cdots -2 B \zeta \textbf{J}\cdot\boldsymbol\ell
\end{equation}

where $\nu_0$ is the vibrational band origin, $B$ is the rotational constant, $D$ is the scalar centrifugal distortion constant, and $\zeta$ is the Coriolis coupling constant. 

\begin{figure}
\centering
\includegraphics[totalheight=11.5cm,scale=1]{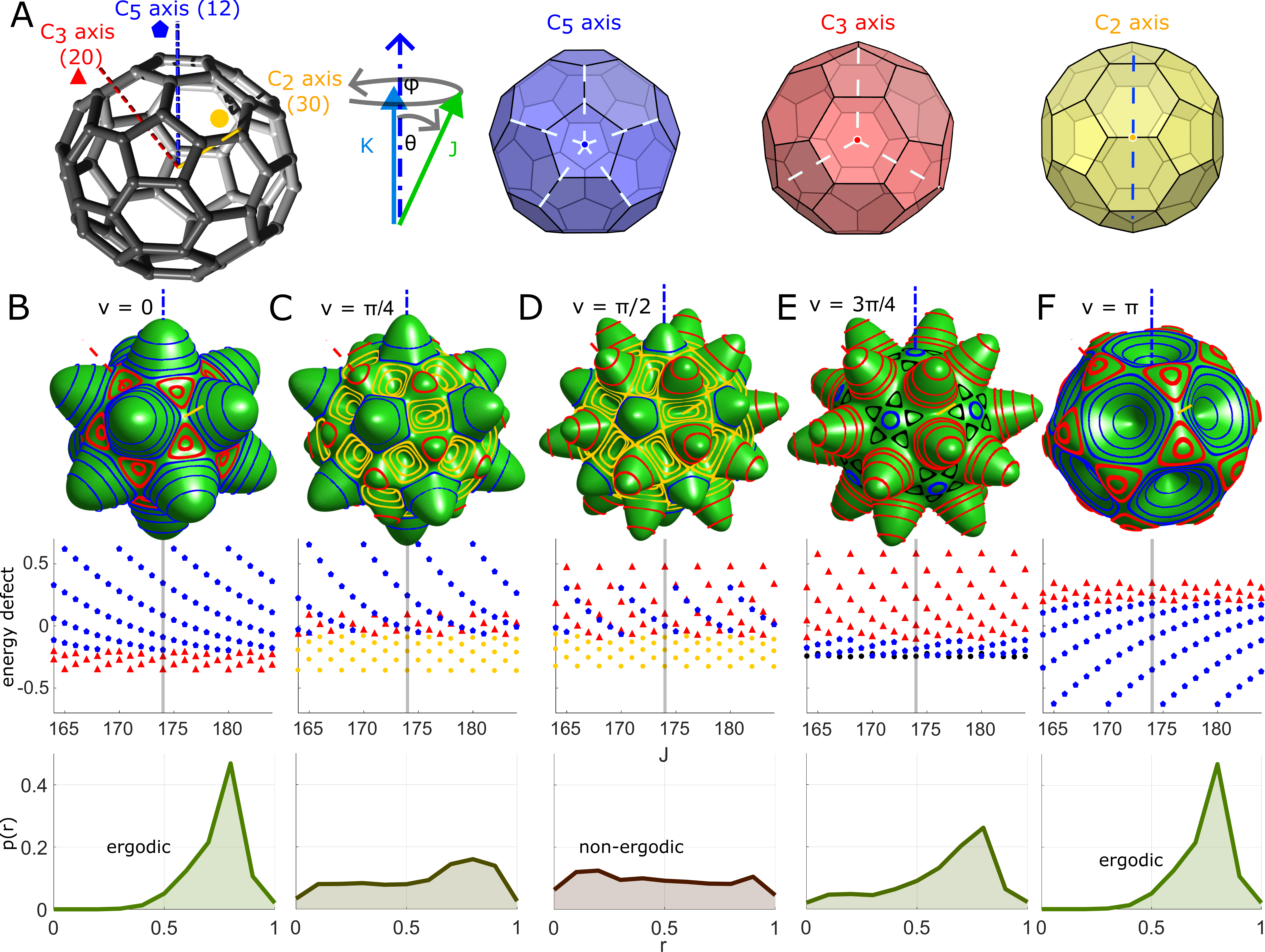}
\caption{\textbf{Rotational energy surfaces and eigenvalues corresponding to icosahedral invariant spherical tensors.} A) Symmetries of C$_{60}$. (L-R): (1) Ball-and-stick model of C$_{60}$, with the three different types of rotational symmetry axes that label stationary points on the rotational energy surface (RES). The degeneracies of the stationary points are listed in parentheses. Color and plot marker coding for each type of rotational symmetry axis are shown. (2) Body-fixed coordinates: polar $\theta$ and azimuthal $\phi$ angles. (3-5) View along $C_5$, $C_3$, and $C_2$ rotational symmetry axes. B)-F) (Top panels) RES, defined by their radii $r(\theta,\phi) = 1+H^{(6+10)}_\text{tensor}(\nu;\theta,\phi)/2$, for $\nu$ ranging from 0 to $\pi$. % Comparing panels B and D shows that higher rank icosahedral harmonics are characterized by finer angular features (more nodes) while preserving icosahedral symmetry. 
Eigenvalues of $1+H^{(6+10)}_\text{tensor}(\nu)/2$ calculated for the fully symmetrized $J=174$ subspace are plotted on the surface as radial contours, colored corresponding to their dominant rotational symmetry character. %Black contours correspond to totally symmetric (`A') eigenstates appropriate for $^{12}$C$_{60}$~\cite{SI}. 
%As $\nu$ varies, the C$_5$, C$_3$, or C$_2$ eigenvalues ``rotate" through each other. 
(Center panels) Energy defects (eigenvalues of $H^{(6+10)}_\text{tensor}(\nu)$) over a range of $J$. The gray vertical line highlights $J=174$. Eigenvalues are plotted using the marker corresponding to their dominant C$_5$, C$_3$, or C$_2$ character (panel A). Near the separatrices, the assignment is somewhat ambiguous due to strong mixing. (Bottom panels) Distribution $p(r)$ of energy gap ratios $r$ (see text) calculated from energy defect spectra aggregated over $J=0-400$. Away from $\nu = 0,\pi$ (panels C-E) the finite value of $p(r)$ as $r\to 0$ is a signature of ergodicity breaking.
%Finally, the character of the symmetry axes is indicated as + (maximum), - (minimum), or 0 (saddle point) in Panels B), D), and E).
}
	\label{fig:tensor_intro}
\end{figure}

Ro-vibrational fine structure is encoded in the tensor Hamiltonian $H_{\mathrm{tensor}}$. For simplicity we consider a pure rotational tensor Hamiltonian 
consisting of the two lowest-order ``icosahedral invariants". These are linear combinations of spherical tensors of the same rank that transform according to the totally symmetric irreducible representation in the icosahedral point group (I$_h$)~\cite{SI}. They can be expressed~\cite{Zheng1996} in the basis of spherical harmonics $Y^k_q(\theta,\phi)$ of degree $k$ and order $q$, which depend explicitly on the molecular frame's polar $\theta$ and azimuthal $\phi$ angles (Figure~\ref{fig:tensor_intro}A). The first two nontrivial (anisotropic) icosahedral invariants, with ranks 6 and 10, are given by:
\begin{align}
    T^{[6]}(\theta,\phi) &= \frac{\sqrt{11}}{5} Y^6_0(\theta,\phi) + \frac{\sqrt{7}}{5} (Y^6_5(\theta,\phi) -Y^6_{-5}(\theta,\phi)) \label{eq:sph_harm_6} \\ 
    \notag
        T^{[10]}(\theta,\phi) &= \frac{\sqrt{3\cdot 13\cdot 19}}{75} Y^{10}_0(\theta,\phi) - \frac{\sqrt{11\cdot 19}}{25} (Y^{10}_5(\theta,\phi) -Y^{10}_{-5}(\theta,\phi))\\ 
        &+ \frac{\sqrt{3\cdot 11\cdot 17}}{75} (Y^{10}_{10}(\theta,\phi) + Y^{10}_{-10}(\theta,\phi))\label{eq:sph_harm_10} 
\end{align}
from which we construct a truncated tensor Hamiltonian
\begin{align}\label{eq:H_tensor}
    H_\text{tensor}^\text{(6+10)} =\gamma ( \cos\nu T^{[6]} + \sin\nu T^{[10]} ).
\end{align}
This Hamiltonian is parametrized by an overall scaling factor $\gamma$ and mixing angle $\nu$ such that $\nu=0$ and $\nu=\pi/2$ correspond to pure $T^{[6]}$ and pure $T^{[10]}$, respectively. 

All operators that are incompatible with icosahedral point group symmetry, including any spherical harmonics of rank 1-5, 7-9, 11, 13, 14, ...~\cite{Bunker1980} vanish from the Hamiltonian. The use of full ro-vibrational tensor operators is unlikely to change the picture qualitatively, particularly when $J$ (approximately 100-300) is much greater than the vibrational angular momentum quantum number $\ell=1$~\cite{SI}. These polyhedral invariants are similar to those used to describe the crystal field splitting of electronic orbitals due to an external lattice environment~\cite{Lea1962} or the ligand field splitting in transition metal complexes~\cite{Griffith1935}.

The energetic correction, or \emph{energy defect}, associated with orienting \textbf{J} in different directions in the molecule frame, can be visualized by the altitude of a semi-classical ``rotational energy surface" (RES), defined at a fixed $J$.
Various possible icosahedral RES's, defined by their radii $r(\theta,\phi) = 1+H^{(6+10)}_\text{tensor}(\nu;\theta,\phi)/2$, corresponding to different mixing angles $\nu$, are plotted for $J=174$ in the top panels of Figures~\ref{fig:tensor_intro}B-F. Stationary points always lie on C$_2$, C$_3$, or C$_5$ rotational symmetry axes. However, as $\nu$ varies, they change in character between minima, maxima, and saddle points. The RES dictates the dynamics of \textbf{J} in the molecule frame~\cite{Harter1978,Harter1981,Harter1984a,Harter1988,Sadovskii1990}, analogous to how an adiabatic potential energy surface steers the relative motions of nuclei~\cite{Khauss1970}. During free evolution, the trajectory of \textbf{J} follows an equipotential contour of the RES. 

In a full quantum mechanical treatment, the perturbation $H^{(6+10)}_\text{tensor}$ leads only to discrete energy defects $e_i$ given by the eigenvalues of $H^{(6+10)}_\text{tensor}$ in a fully symmetrized fixed-$J$ subspace. These orbits trace out the closed contours on the RES in Figure~\ref{fig:tensor_intro}. The orbits may also be obtained directly from the RES: they are the trajectories which both (i) satisfy a Bohr quantization condition~\cite{Harter1984a,Harter1989}, and (ii) transform according to the totally symmetric irreducible representation of the icosahedral point group~\cite{SI}. The latter condition accounts for the quantum indistinguishability of each bosonic nucleus in the $^{12}$C$_{60}$ isotopologue~\cite{Johnson1990,Changala2019}, and is analogous to the selection of odd or even rotational states in ortho- or para-hydrogen molecules, respectively.  
The energy defects are plotted for a range of $J$ in the center panels of Figures~\ref{fig:tensor_intro}B-F and may be expected to appear in the ro-vibrational fine structure of $^{12}$C$_{60}$. 

In the preceding discussions, we have focused solely on the geometric effects of icosahedral symmetry. In general, however, the measured tensor defect spectra may exhibit additional $J$-dependent scaling and offsets, which depend on intra-molecular couplings specific to $^{12}$C$_{60}$. To extract all of these features, we now turn to the experimental measurements.

\textit{Methods-Experimental.}
To resolve ro-vibrational perturbations in $^{12}$C$_{60}$, in this work we explored the P-branch region of the 1185~cm$^{-1}$ T$_{1u}(3)$ band, first identified as a region of potential interest in Ref.~\cite{Changala2019}. Using cavity-enhanced continuous-wave (cw) spectroscopy with a quantum cascade laser (QCL) source, we achieved a minimum absorption sensitivity $\alpha_{min} = 2.1\times 10^{-10}~\textrm{cm}^{-1}\textrm{Hz}^{-1/2}$, or 1,000-fold better detection sensitivity per spectral element than in Ref~\cite{Changala2019} ($\alpha_{min} = 2.2\times 10^{-7}~\textrm{cm}^{-1}\textrm{Hz}^{-1/2}$ per comb mode).  600~MHz-wide absorption spectra were acquired by simultaneously scanning the QCL frequency and free spectral range of the enhancement cavity across molecular absorption lines, and recording the frequency-dependent absorption. These spectra were stitched together by a combination of direct calibration of the QCL frequency with a Fourier transform spectrometer, and comparison to matching spectral features in the broadband, low signal-to-noise (SNR) frequency comb spectrum of Ref~\cite{Changala2019}. We obtain an absolute frequency accuracy of $\sim6~$MHz throughout the entire measured frequency range, limited by the resolution of the reference frequency comb spectrum. Finally, to ensure consistency of the absorption signal over multiple days of data-taking, we have periodically re-measured the molecular absorption feature at R($J\approx180$) at $\nu=1186.27$~cm$^{-1}$ and normalized all data taken around the same time to its line strength and measured cavity finesse. 

\begin{figure}
\includegraphics[width=\columnwidth,scale=1]{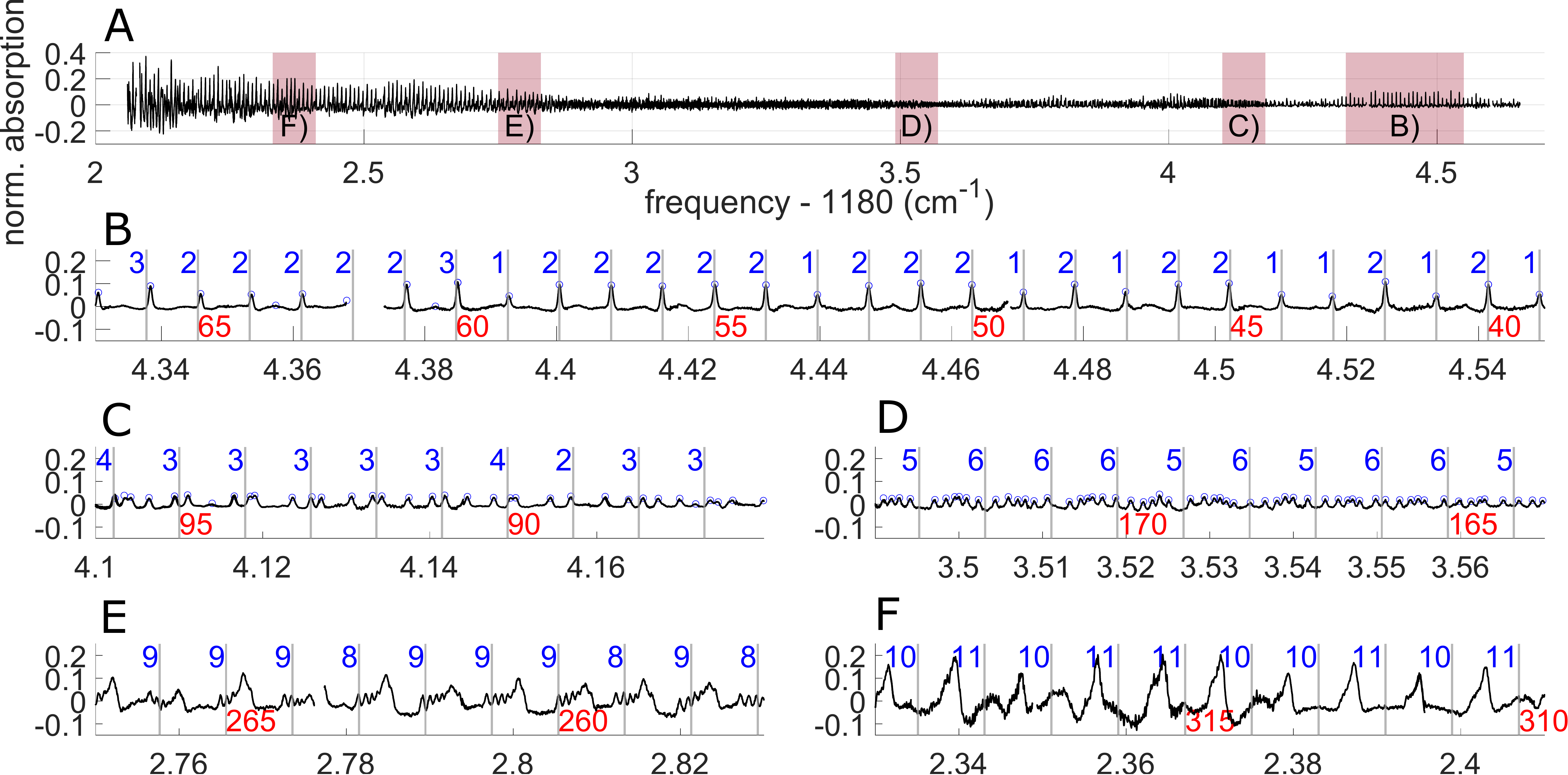}
\caption{\textbf{Direct continuous-wave (cw) absorption spectroscopy of C$_{60}$ P-branch.} A) Complete normalized cw spectrum of P-branch. % Raw absorption has been multiplied by $(2J+1)^{-1}e^{B''J(J+1)/k_BT}$ %(blue dashed line)
in order to expose the J-dependence of the nuclear spin weights. Red highlighted regions are shown in greater detail in subsequent panels. B)-F) Zoom into red highlighted regions. B) Zoom into low-J region of P-branch. This relatively unperturbed region of the P-branch is fit to a rigid-rotor Hamiltonian to obtain the  rotational spacing $2B(1-\zeta)$. Fitted line positions are indicated by the grey vertical grid lines, with $J$ labelled in red. Blue numbers denote calculated nuclear spin weights, which match the measured peak intensities well. In C) and D), peaks are resolvable and marked with blue circles. In E) and F), spectral congestion prohibits identification of individual peaks. }
	\label{fig:QCL_raw_spectrum}
\end{figure}

\textit{Methods - Analysis.} The culmination of these efforts is the infrared spectrum in Figure~\ref{fig:QCL_raw_spectrum}, spanning the spectral region from  1182.0-1184.7~cm$^{-1}$. At $J\lesssim70$, there is a regular progression of rotational lines, similar to those in the R-branch~\cite{Changala2019}. They rapidly split into intricate patterns before merging at and beyond $J\approx 300$. Zooming into the region labelled B), the rotational line centers can be fit to the scalar part of Equation~\ref{eq:H_tot}, as was done in Ref~\cite{Changala2019}.
\begin{align}
    \nu^P(J)=\nu_0 + B'(J-1)(J+2\zeta)-B'' J(J+1) \label{eq:scalar_ham}
\end{align}
where $J$ here refers to the \textit{ground-state} total angular momentum. The scalar centrifugal distortion term is not significant at our spectral precision and range of $J$. %To be consistent with the spectroscopic constants measured from the R-branch in Ref~\cite{Changala2019}, we have chosen values of %$R = J+1$ is the pure rotational angular momentum in the P-type T$_{1u}^{(+)}$ excited state, 
The fit yields $B''$ = 0.0028~cm$^{-1}$ for the ground state rotational constant, $B' =B'' -2.876\times10^{-7}$~cm$^{-1}$ for the excited state rotational constant, %T$_{1u}^{(+)}$ rotational constant, 
$\zeta=-0.37538$ for the Coriolis coupling constant, %$\ell=1$ is the vibrational angular momentum, 
and $\nu_0=1184.85~$cm$^{-1}$. Equation~\ref{eq:scalar_ham} yields a progression of rotational lines with a spacing of approximately $2B(1-\zeta)$, where $B= (B' +B'')/2$. Note that the spectroscopic constants are under-determined and only serve to facilitate $J$-assignment of the peaks in a manner consistent with the R-branch assignments of Ref~\cite{Changala2019}. %Also, $\zeta = 0.988~\zeta_R$, where $\zeta_R$ corresponds to the R-branch measured in Ref~\cite{Changala2019}. This could be due to scalar coupling of a dark vibrational state to one of T$_{1u}^{(+/-)}$(3), which could modify its effective rotational constant compared to T$_{1u}^{(-/+)}$(3). 
The extrapolated P-branch spectral line positions based on this scalar fit are plotted as gray vertical lines in Figure~\ref{fig:QCL_raw_spectrum} B-F, with every fifth $J$ value labelled in red. The agreement with the measured line positions is excellent in the region $J<70$ where the spectrum appears unperturbed (Figure~\ref{fig:QCL_raw_spectrum}B).

To confirm our $J$ assignment, we compare the peak absorption cross sections to the nuclear spin weights of the ground state rotational levels % In this region, the $2J+1$ degeneracy of the angular momentum lab-frame projections varies slowly, while the nuclear spin weights, labelled in blue, jump non-monotonically between values of 1, 2, and 3. 
and find that they match extremely well. %At large J, the nuclear spin weight approaches $\approx(2J+1)/60$.
Finally, we apply a frequency-dependent scaling factor to the raw absorption spectrum $(2J+1)^{-1}e^{B'' J (J+1)/ k_BT}$. %(plotted as blue dashed line in Figure~\ref{fig:QCL_raw_spectrum}A). 
This removes the effects of lab frame angular momentum degeneracy and the thermal ensemble, emphasizing the dynamics in the molecule fixed frame. %. This rescaling ensures that the absorption cross section due to a single ro-vibrational fine structure component appears constant throughout the P-branch spectrum, exposing the J-dependence of the nuclear spin weights and 

The peaks are identified manually and circled in blue. Figure~\ref{fig:QCL_raw_spectrum}C) and D) show two representative regions, at $J\sim90$ and $J\sim170$, respectively, where peaks can still be individually resolved. The local peak density again matches the predicted nuclear spin weights (in blue), confirming that the ro-vibrational fine structure splitting originates from icosahedral tensor interactions of Equation~\ref{eq:H_tensor} that lift $K$-degeneracy. % of the type in Equation~\ref{eq:wigner_eckart_thm}, which lift K-degeneracy. 
In Figures~\ref{fig:QCL_raw_spectrum}E) and F), we plot two regions where the peaks have begun to merge, and individual peaks can no longer be easily identified ($J>247$). 

\begin{figure}
\centering
\includegraphics[totalheight=13cm,scale=1]
{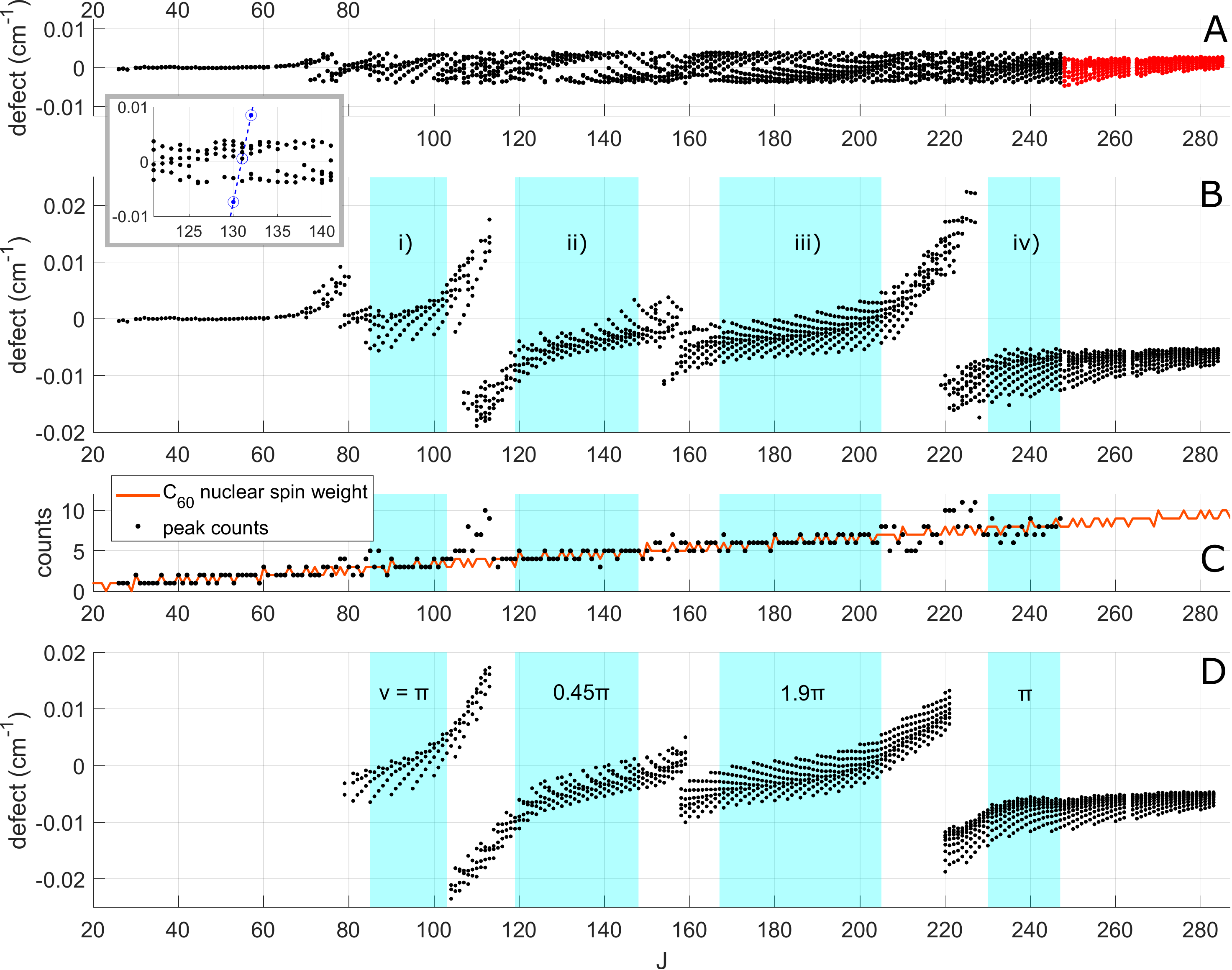}
\caption{\textbf{Obtaining P-branch rotational energy defects vs. $J$. } A) Defect plot from nearest neighbor assignment. Peaks are assigned to nearest $J$ according to rigid rotor model (gray vertical grid lines in Figure~\ref{fig:QCL_raw_spectrum}B-F). Red defect points ($J>247$) correspond to peaks in the absorption spectrum that are no longer individually resolved. Their peak centers are obtained from fitting to a cluster of Voigt lineshapes~\cite{SI}. (Inset) unwrapping procedure follows a series of allowed “moves” which simultaneously translate points in vertical steps of $\pm2B(1-\zeta$) and horizontal steps of $\pm1$. B) ``Unwrapped" defect plot. Four avoided crossings with varying strengths are seen at $J\simeq$ 80, 110, 160, and 220. Defect patterns resemble eigenvalue spectra of icosahedral invariant tensors.  C) Calculated nuclear spin weights overlaid with measured peak counts. Blue highlighted sections show excellent agreement between calculated nuclear spin weights of C$_{60}$ and assigned peak counts. D) Reconstruction of perturbed sections of P-branch spectrum based on eigenvalue spectra of mixed tensor operator $H^{(6+10)}_\text{tensor}(\nu)$. Four perturbed portions of panel B) are reproduced with four mixing angles $\nu$. %The nature of the $C_5$, $C_3$, and $C_2$ axes is also indicated, following the same convention as in Figure~\ref{fig:tensor_intro}.
}
	\label{fig:defect_plot}
\end{figure}

%Having applied initial processing to a raw one-dimensional absorption spectrum, our next objective is to expose the J-dependence of the energy splitting. 
Interpreting the tensor splittings requires assigning each absorption peak to a particular $J$. %This is complicated by the fact that the energy splittings may exceed the $2B(1-\zeta)$ rotational spacing and ``wrap" to adjacent $J$'s. 
We begin by assigning each peak to its nearest $J$ value according to the scalar fit of Equation~\ref{eq:scalar_ham}. Subtracting the scalar contribution (Equation~\ref{eq:scalar_ham}) from the central frequency of each peak generates a single ``period" of a Loomis-Wood-like defect plot in Figure~\ref{fig:defect_plot}A). There remains some ambiguity in the defect assignments, as illustrated in the inset of Figure~\ref{fig:defect_plot}A: each defect is constrained to the line with slope  $2B(1-\zeta)$ which passes through its current position.

By carefully rearranging individual defects according to these discrete allowed ``moves", we unwrap five distinct regions exhibiting continuous-looking patterns (Figure~\ref{fig:defect_plot}B). Because of the discontinuities at $J\simeq$ 80, 110, 160, and 220, there was still some ambiguity in the overall shift of the four perturbed sections labelled i)-iv). To remove this ambiguity, we recognized %that each of the perturbed sections, labelled i)-iv) in Figure~\ref{fig:defect_plot}B), have 
each section's dominant pure-rank tensor character as follows: i) $-T^{[6]}$; ii) $T^{[10]}$; iii) $T^{[6]}$; iv) $-T^{[6]}$. Eigenvalue spectra in Figures~\ref{fig:tensor_intro}B), D), and F) show that the extremal eigenvalues associated with $C_n$ rotational symmetry axes occur when $J$ is an integer multiple of $n$. The $J$-assignment depicted in Figure~\ref{fig:defect_plot}B is one that satisfies this condition for all sections simultaneously. This final $J$-assignment is confirmed by the excellent agreement between $J$-resolved peak counts and calculated nuclear spin weight far from the discontinuities Figure~\ref{fig:defect_plot}C). %confirms the existence of perturbative spectral shifts in the vicinity of $J\approx$ 80, 110, 160, and 220, which will be discussed further below. 

\textit{Results and Discussion.} Having obtained a satisfactory $J$-assignment of measured defects from the raw absorption spectrum, we now turn to its interpretation. We parametrize each of the four perturbed regions i)-iv) in Figure~\ref{fig:defect_plot}B) in terms of a mixed tensor (Equation~\ref{eq:H_tensor}), $J$-dependent scaling $b(J)$, and $J$-dependent scalar offset $a(J)$:
\begin{equation}\label{eq:fitting_fn}
    a(J) + b(J) \times e(\nu;J,K)
\end{equation}
where $e(\nu;J,K)$ are the $J$-dependent tensor splittings as plotted in the lower panels of Figures~\ref{fig:tensor_intro}B-F.

First, $a(J)$ is obtained from the observed mean defect of each section. Next, by performing a point-cloud registration~\cite{Yang1992,Besl1992,Segal2010} to the theoretical eigenvalue spectra and the measured defect plot, we assign a best-fit mixing angle $\nu$ to each section: i) $\pi$; ii) 0.45$\pi$; iii) 1.9$\pi$, and iv) $\pi$~\cite{SI}. Finally, $b(J)$ is obtained from a least-squares fit to a polynomial in $J$~\cite{SI}. %Finally, by minimizing the rms error, good agreement was found with $b(J)/(10^{-3}$~cm$^{-1})\approx~$ i) 6.41; ii) 7.13; iii) 7.15; and iv) $8.2+4.9\times 10^{-2} J -2.4\times 10^{-4} J^2$~\cite{SI}. %10.0
The resulting reconstructed defect plot for regions i)-iv) is plotted in Figure~\ref{fig:defect_plot}D. 

The abrupt changes in mixing angles $\nu$ are associated with transitions between ergodic and non-ergodic rotational dynamics as the molecule ``spins up" to higher $J$. We find that these dynamics transition from ergodic for region i), to non-ergodic for region ii), and back again for regions iv) and v). 

The origin of this ergodicity breaking can be understood semi-classically using the RES's in Figure~\ref{fig:tensor_intro} B), D), and F). The dynamics of $\mathbf{J}$ are ergodic when time evolution explores the full space of symmetry-allowed states at the same energy. For region iii), the dominant tensor character is $T^{[6]}$. Naively, the existence of 12 disconnected trajectories encircling the C$_5$ axes breaks ergodicity. However, these trajectories cannot be distinguished in $^{12}$C$_{60}$: due to the indistinguishability of the $^{12}$C nuclei, all 12 semi-classical trajectories correspond to a single quantum state given by their fully symmetric superposition. As such, the $\mathbf{J}$ dynamics \emph{do} explore the full range of states at a given energy, and hence is ergodic. The same argument applies for regions i) and iv), which differ from iii) only in the sign of energy defects (Figure~\ref{fig:tensor_intro}F).

In contrast, for region ii) the dominant tensor character is $T^{[10]}$. The C$_5$ and C$_3$ axes both correspond to peaks on the RES, and host trajectories over the same range of energies (Figure~\ref{fig:defect_plot}D). Trajectories encircling the C$_5$ and C$_3$ axes \emph{are} distinguishable, and hence correspond to distinct quantum states. The quantum tunnelling between these trajectories is unable to restore the ergodicity: the tunnelling integral between C$_5$ and C$_3$ is exponentially small in $J$~\cite{Auerbach1994,vanHemmen1986}, whereas the level spacing only scales as $1/J$. The scaling of the tunnelling integral follows from standard WKB results~\cite{Landau1958}, while the level spacing can be seen from comparing the relatively fixed bandwidth of energy defects~(Figure~\ref{fig:defect_plot}B) with the nuclear spin weight $\sim(2J+1)/60$. Consequently, for our measured $J$ (well within the large-$J$ limit), tunnelling corrections are typically only perturbative.

Level statistics provide a simple probe of this ergodicity breaking in molecular spectra~\cite{Sundberg1985,Zimmermann1988}. Ergodic and non-ergodic dynamics are respectively associated with level repulsion and its absence~\cite{Wigner1967}. 
A useful diagnostic tool is the distribution $p(r)$, where $r$ is the ratio of consecutive level spacings $e_i$~\cite{Oganesyan2007,Atas2013}:

%To see this, we examine the statistics of energy gap ratios:
\begin{align}
    r_i &= \text{min}\left(\frac{s_i}{s_{i-1}},\frac{s_{i-1}}{s_{i}}\right)\\
    s_i &=e_{i+1}-e_i
\end{align}

%where $e_i$ are consecutive energy levels. %The behaviour of the distribution $p(r)$ as $r\rightarrow 0$ characterizes the ergodicity of a many-body spectrum: 
In the limit of $r \to 0$, level repulsion in an ergodic system causes $p(r)\to 0$, while for a non-ergodic system $p(r)\to \text{const}$. Similar energy level statistics have been used to analyze systems as diverse as nuclear spectra~\cite{Wigner1967} and ultra-cold atomic scattering~\cite{Frisch2014}, and many-body localization~\cite{Abanin2019}.

\begin{figure}
\centering
\includegraphics[width=\textwidth]{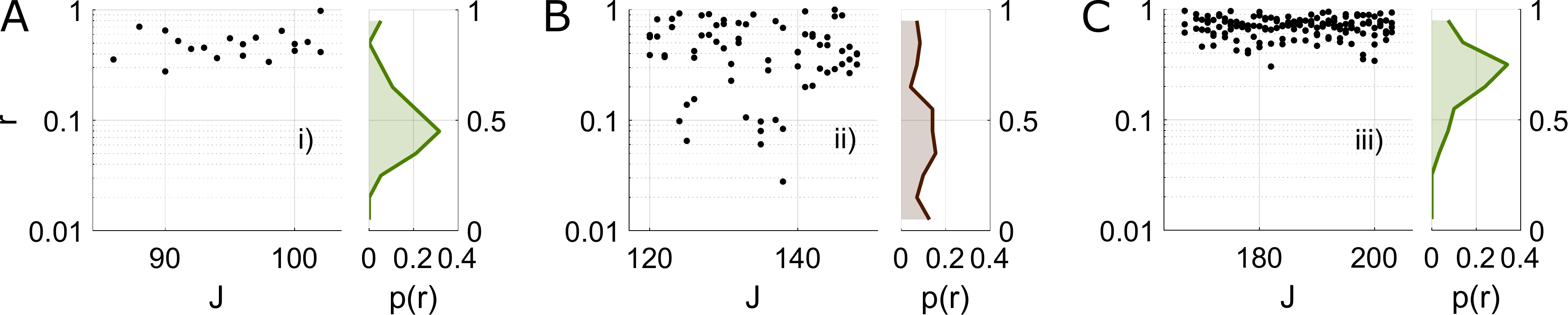}
\caption{\textbf{Energy level statistics in ergodic and non-ergodic regimes.} A)-C) (Left panels) Gap ratio $r$ as a function of $J$ calculated from sections i)-iv) of the defect spectrum (Figure~\ref{fig:defect_plot}B). Gap ratios are only plotted at $J$ values for which the peak counts match the calculated nuclear spin weight (Figure~\ref{fig:defect_plot}C). (Right panels) Normalized distribution $p(r)$ of gap ratios aggregated over sections i)-iii). Note the change from logarithmic to linear $r$ scale between left and right panels. Sections i) and iii) exhibit level repulsion, a signature of ergodicity, whereas section ii) does not, indicating non-ergodicity. }
\label{fig:gap_ratio}
\end{figure}

Figure~\ref{fig:gap_ratio}A)-C) show the energy gap ratios computed from sections i)-iii), respectively, of the experimental defect plot (Figure~\ref{fig:defect_plot}B). Here, sections i) and iii) exhibit significant level repulsion characteristic of ergodicity, whereas section ii) does not, indicating non-ergodicity. We aggregate the energy gap ratios over each one of sections i)-iii) and their respective distributions in Figure~\ref{fig:gap_ratio}D)-F). These distributions confirm the presence of level repulsion in sections i) and iii), and its absence in section ii). %The distributions show the expected behaviour as $r\rightarrow 0$ for ergodic and localized regimes. 

We now turn from the dynamical interpretation of ro-vibrational defects to discuss their physical origin. Further analysis of Figure~\ref{fig:defect_plot}B reveals the defects arise from ro-vibrational coupling between the bright P-type excited T$_{1u}^{(+)}(3)$ state and a background of perturbing dark states. Specifically, both the discontinuities and excess observed peaks at specific $J$ values are characteristic of avoided crossings with perturbing zero-order dark vibrational combination states. % into which the T$_{1u}(3)$ state is embedded~\cite{SI}. 
As they cross T$_{1u}^{(+)}(3)$ from below, ro-vibrational coupling lifts the degeneracy of $J$ multiplets in the T$_{1u}^{(+)}(3)$ state, imparting tensor character that depends on the identity of the perturbing vibrational state. 
%these dark states transfer some of their individual tensor character onto the T$_{1u}(3)$ state via ro-vibrational coupling. S
The tensor character of the T$_{1u}^{(+)}(3)$ state (specifically the fitted $\nu$ values) is stable in between avoided crossings, suggesting each of these sections is dominantly affected by just one perturbing state at a time. At the avoided crossings, this is expected to break down, as made particularly evident by the rapid change in mixing angle just before and after the avoided crossing at $J=160$ of Figure~\ref{fig:defect_plot}B. We note there is no apparent structure to the changes in mixing angle and coupling strength induced by the different avoided crossings, suggesting that the perturbing dark states are distinct, and not part of the same Coriolis manifold. Finally, using the observed density of perturbing states $\rho_\textrm{obs}\approx2/$cm$^{-1}$ and average measured vibrational coupling strength (bandwidth of the avoided crossings) of $\beta_\textrm{avg}\approx2\times10^{-2}$~cm$^{-1}$~\cite{SI}, we arrive at an IVR threshold parameter~\cite{Logan1990} $T(E)=\sqrt{2\pi/3}~\rho_\textrm{obs}\beta_\textrm{avg} \approx0.06\ll1$. Our $^{12}$C$_{60}$ rotational ergodicity transitions are observed well below the IVR threshold, a conclusion supported by the spectroscopically well-resolved $T_{1u}(3)$ band. 

%with no simple relation to one another.

%Finally, the abrupt and apparently random change of the mixing angle and coupling strengths at the avoided crossings suggest that these perturbing states are totally distinct with no simple relation to one another. 
\textit{Conclusion.} 
For the first time, we have measured and characterized icosahedral tensor ro-vibrational coupling in $^{12}$C$_{60}$. %, implying a class of emergent behaviour characteristic of rotating finite quantum systems~\cite{Yannouleas2007,Birman2013}. 
Analysis of the spectrum of ro-vibrational defects reveals that as the molecule is spun up to higher $J$, there is a series of transitions in the dynamical behaviour of \textbf{J} in the fixed body frame. Specifically, \textbf{J} switches between ergodic and non-ergodic behaviour at particular $J$ values, leaving a characteristic imprint on the defect level statistics.
%The spectrum of ro-vibrational defects indicates that the precession of \textbf{J} about various molecular symmetry axes undergoes transitions between ergodic and non-ergodic regimes at multiple $J$ values. 
These ergodicity transitions arise from dark vibrational combination states which cross the T$_{1u}^{(+)}(3)$ state at particular $J$ values, transferring their anisotropic character onto the T$_{1u}^{(+)}(3)$ state via ro-vibrational coupling.

Our measurements open the door to a rich hierarchy of emergent behaviour in C$_{60}$ isotopologues, accessible at ever higher spectral resolution. 
%pins the molecule to precess about just one of several equivalent symmetry axes, thereby 
The small nuclear spin-rotation interaction, for example in $^{13}$C substituted isotopologues of C$_{60}$, can have a magnified effect due to the small superfine splittings near RES extrema. Such “hyperfine” coupling can lead to spontaneous symmetry breaking in a finite system~\cite{Harter1979,Borde1980}. These insights could prove useful for exploiting the exotic orientation state space of C$_{60}$ for quantum information processing~\cite{Albert2020} and for investigating the quantum to classical transition of information spreading~\cite{Zhang2022}. Ultimately, spectroscopy of C$_{60}$ isotopologues at ever higher spectral resolution promises to uncover deeper insights into the emergent dynamics of mesoscopic quantum many body systems.

\clearpage
\bibliography{refs,SI,bib2}

\bibliographystyle{Science}

\section*{Acknowledgments:}
This research was supported by AFOSR grant no. FA9550-19-1-0148; the
National Science Foundation Quantum Leap Challenge Institutes (grant QLCI OMA-2016244); 
the National
Science Foundation (grant Phys-1734006); the U.S. Department of Energy, Office of Science, National Quantum Information Science Research Centers, Quantum Systems Accelerator (QSA); and the National Institute of Standards and Technology. D.R. acknowledges support from the Israeli council for higher education quantum science fellowship, and is an awardee of the Weizmann Institute of Science – Israel National Postdoctoral Award Program for Advancing Women in Science. P.J.D.C. and N.Y.Y. acknowledge support from the AFOSR MURI program (FA9550-21-1-0069). T.T. acknowledges support from the NSF EPSCoR RII Track-4 Fellowship No. 1929190. We gratefully acknowledge comments on the manuscript from Aaron W. Young and Ya-Chu Chan. 

\textbf{Author Contributions: }L.R.L, D. R., and J.Y. designed, discussed, and performed the experiment and analyzed the data. All authors participated in theory discussions and calculations, and writing of the paper. 

\textbf{Competing interests:} The authors declare no competing interests. 

\textbf{Data Materials and Availability: }All data are available in the main text, in the
supplementary information, or through Harvard Dataverse. All materials requests and correspondence should be directed to Lee R. Liu or Jun Ye.

%Here you should list the contents of your Supplementary Materials -- below is an example. 
%You should include a list of Supplementary figures, Tables, and any references that appear only in the SM. 
%Note that the reference numbering continues from the main text to the SM.
% In the example below, Refs. 4-10 were cited only in the SM.     
\clearpage
%$H^{(6+10)}_\text{tensor}(\nu)=-H^{(6+10)}_\text{tensor}(\nu+\pi)$, so we need only display RES for the interval $0\leq\nu\leq\pi$.
% RES for $H^{(6+10)}_\text{tensor}(\nu)$
%The RES is defined by its radius  $r(\theta,\phi) = 1+H^{(6+10)}_\text{tensor}(\nu;\theta,\phi)/2$ with a scaling and ``spherical offset" for clarity. The radial coordinate $r$ of the RES then represents the ro-vibrational fine structure energy (scaled by 1/2 and offset by 1). 

\clearpage
\newcommand{\beginsupplementEqs}{%
        \setcounter{equation}{0}
        \renewcommand{\theequation}{S\arabic{equation}}%
     }
\beginsupplementEqs

\section*{Supplementary Information}
%\section*{Supplementary materials}
Materials and Methods\\
Supplementary Text\\
Tables S1 to S3\\
Figs. S1 to S4\\
References \textit{(28-50)}

\paragraph*{Materials and Methods.}

\paragraph*{Icosahedral invariant tensor operators}\label{sec:tensor_analysis}
The full tensor part of the molecular Hamiltonian, including rotational and vibrational degrees of freedom, is given by
\begin{align}\notag
    H_\text{tensor} &= \sum_{r_1,r_2,r_3} t_{r_1,r_2,r_3}T^{[r_3]}[r_1,r_2] \\
    & = \sum_{r_1,r_2} \{t_{r_1,r_2,6} T^{[6]}[r_1,r_2]+t_{r_1,r_2,10} T^{[10]}[r_1,r_2] +\cdots \}
\end{align}
where $r_1$ is the rank of the vibrational angular momentum operator, $r_2$ is the rank of the total angular momentum operator, and $r_1+r_2=r_3$. The correlation between SO(3) and $I_h$ restricts the sum to $r_3 = 6,10,12,\cdots$ (the terms with $r_3=0$ appear in the scalar part of the Hamiltonian). This is related to the allowed rotational states $J = 0,6,10,\cdots$ 
For simplicity, we will consider a pure rotational tensor Hamiltonian, i.e, which does not depend on vibrational angular momentum. %This will be justified further below. 
\begin{equation}
    H_\text{tensor}^0 =t_{0,6,6} T^{[6]}[0,6]+t_{0,10,10} T^{[10]}[0,10] +\cdots
\end{equation}

Equation~\ref{eq:fitting_fn} provides some justification for why we have considered only pure rotational tensors; since we expect similar patterns of icosahedral splittings to appear in pure rotational or ro-vibrational manifolds~\cite{Harter1989}, and the vibrational quantum number is constant throughout the excited state T$_{1u}(3)$ manifold, we can absorb any vibrational dependence into the $J$-dependent scaling factor $b(J)$ (discussed more below).

\paragraph*{Eigenvalues of $H_\text{tensor}$ and ``A"-symmetry selection.}\label{sec:A_symm_selection}

Here, we describe the procedure of calculating the eigenvalues of the mixed 6-th and 10-th rank  tensor  Hamiltonian  (Equation~\ref{eq:H_tensor}) \cite{Harter1989}, 
\begin{comment}
    \begin{equation}\label{H_tensor}
\hat{H}_\text{tensor}^\text{(6+10)} = \cos\nu T^{[6]} + \sin\nu T^{[10]} 
\end{equation}
where $\nu$ is the mixing angle, 
\end{comment}
where the 6-th and 10-th order icosahedral tensors are given by
\begin{align}
    T^{[6]} &=  \frac{\sqrt{11}}{5}  T^{(6)}_0  +  \frac{\sqrt{7}}{5} (T^{(6)}_5  +   T^{(6)}_{-5}) \\
 T^{[10]} &=  \frac{\sqrt{3\cdot 13\cdot 19}}{75} T^{(10)}_0  -  \frac{\sqrt{11\cdot 19}}{25} (T^{(10)}_5  -   T^{(10)}_{-5}) +   \frac{\sqrt{3\cdot 11\cdot 17}}{75} (T^{(10)}_{10}  +   T^{(10)}_{-10}) 
 \end{align}

We obtain the matrix elements of these tensors by diagonalizing $T^{[k]}$ in the basis of symmetric top eigenstates  $|J K\rangle=|J K M\rangle$ where $K,M=-J,-J+1,\cdots,+J$ are the body-fixed and space-fixed projections of $J$, respectively ($M$ will be omitted in the following for notational simplicity). According to the Wigner-Eckart theorem, the matrix elements of $T^{[k]}$ can be factorized into a reduced matrix element $\langle J ||T^{(k)}||J\rangle$ which carries the dynamical content of the tensor operator and a component which only depends on geometry, expressed here in terms of Wigner 3-j symbols~\cite{Zare1988}:
\begin{align}\label{H_tensor_mel_6}\notag
\langle JK |   T^{[6]}  |JK'\rangle &=  \langle J || T^{(6)} || J\rangle  (-1)^{J-K} (2J+1)^{1/2}   \\ \notag
\biggl{\{}  &\frac{\sqrt{11}}{5} \threejm{J}{-K}{6}{0}{J}{K'}  
 +  \\&\frac{\sqrt{7}}{5} \left[ \threejm{J}{-K}{6}{5}{J}{K'} 
-  \threejm{J}{-K}{6}{-5}{J}{K'} \right] \biggr{\}}
\end{align}
for the $6^{th}$ rank tensor interaction and
\begin{align}\label{H_tensor_mel_10}\notag
\langle JK |   T^{[10]}  |JK'\rangle &=   \langle J || T^{(10)} || J\rangle  (-1)^{J-K} (2J+1)^{1/2}  \\  \notag
\biggl{\{}  &\frac{\sqrt{3\cdot 13\cdot 19}}{75}  \threejm{J}{-K}{10}{0}{J}{K'}  \\ \notag
-  & \frac{\sqrt{11\cdot 19}}{25}\left[ \threejm{J}{-K}{10}{5}{J}{K'} -  \threejm{J}{-K}{10}{-5}{J}{K'} \right]  \\
+  &\frac{\sqrt{3\cdot 11\cdot 17}}{75} \left[ \threejm{J}{-K}{10}{10}{J}{K'} +  \threejm{J}{-K}{10}{-10}{J}{K'} \right] \biggr{\}}
\end{align}
for the $10^{th}$ rank tensor interaction (note that the {reduced} matrix element used by Harter and Weeks \cite{Harter1989} have been scaled by the factor $(2J+1)^{1/2}$ to bring it into agreement with the convention of Ref.~\citen{Zare1988}). Diagonalization yields $2J+1$ possible eigenvalues, corresponding to the white contours on the RES of Figures~\ref{fig:tensor_intro}B-F. To account for the quantum permutation statistics of the bosonic $^{12}$C nuclei comprising $^{12}$C$_{60}$, we select only the eigenvalues belonging to totally symmetric eigenstates in I$_h$~\cite{SI,Bunker1980}. This reduces the number of fine-structure components, or the ``nuclear spin weight", to approximately $(2J+1)/60$. 

In the following, we will set both of the {reduced} matrix elements to 1 to expose  the mixing  effects, as done by Harter and Weeks \cite{Harter1989}. It is important to keep in mind, however, that these matrix elements {may} scale  differently with  $J$ as $\langle J || T^{(r)} || J\rangle \simeq J^r$, so the  10-th rank tensor  contribution will  become dominant at high $J$.

To select the A symmetry species for  the 6-th rank tensor spectrum of $^{12}$C$_{60}$, we implemented the sequencing wheel in Fig. 3 of Harter and Weeks \cite{Harter1989}. To this end, we  first calculate  $(J)\,\mathrm{mod}(10)$ and locate the highest-energy symmetry species on the outside of the wheel. For instance, for $J=100$ the highest-energy species is A. We then follow the direction indicated by the arrow and count the different symmetry species including their degeneracies until we encounter A again. The degeneracies of the relevant icosahedral symmetry species are 3 for T$_1$ and T$_3$, 4 for G, and 5 for H. Repeating this procedure until the last eigenvalue gives the indices of the 6-rank tensor eigenvalues of A symmetry. For $J=100$, we have 4 A-symmetry eigenvalues shown by the red dots in Fig. 1. For $J\leq 30$ this procedure gives either 0 or 1  in agreement with Ref.~\citen{Bunker1980}.

An important problem now arises of identifying the A symmetry eigenvalues of the mixed (6 + 10)-th rank tensor. Because the order of the species changes as  the 10-th rank tensor is admixed, one can no longer use the sequencing wheel designed specifically for the 6-th order tensor interaction  \cite{Harter1989}. No similar wheels are available  in the literature for mixed 6-th and 10-rank tensors.

We suggest the following solution to  this problem. Because we can identify the A species among the eigenvalues of the  6-rank tensor Hamiltonian using the sequencing wheel \cite{Harter1989}, the corresponding eigenfunctions have the proper A symmetry with respect to the icosahedral group operations.
Thus, we can use  these eigenfunctions as a new basis to construct the matrix of the (6 + 10)-th rank tensor interaction. Because the full Hamiltonian is block-diagonal in the symmetry-adapted basis, diagonalization of this matrix will give the rigorous A-symmetry eigenvalues of the mixed (6 + 10)-th rank tensor Hamiltonian. 
A more conventional  approach would be to use the properly symmetrized $\ket{JK}$ basis functions to construct the Hamiltonian matrix. Constructing such icosahedral harmonics is a non-trivial  task, but efficient algorithms are known for generating A-symmetry harmonics \cite{Zheng1996}.

To implement our approach, we first select the 6-th rank A symmetry species at $\nu=0$ shown by the dots in Figure~\ref{fig:SI_A_selection} using the sequencing wheel approach \cite{Harter1989}. We then transform the $(2J+1)\times (2J+1)$ Hamiltonian matrix $\mathbf{H}$ to the A-symmetry basis
\begin{equation}\label{H_A}
\mathbf{H}_A = \mathbf{C}^T \mathbf{H}\mathbf{C},
\end{equation}
where $\mathbf{C}$ is a rectangular matrix whose columns are the preselected eigenvectors of A-symmetry. The resulting matrix $\mathbf{H}_A$ is a small matrix  ($4\times 4$ for $J=100$) containing only the A symmetry species. Note that $\mathbf{H}_A$ is diagonal for $\nu=0$ and close to diagonal for small $\nu$.

Figure~\ref{fig:SI_A_selection} shows the eigenvalues of the full mixed-rank tensor Hamiltonian  for $J=100$ as a function of the mixing angle $\nu$. We observe complex crossing and anti-crossing patterns consistent with Fig. 6 of Harter and Weeks \cite{Harter1989}.
The 6-th rank A-symmetry level pattern  is  preserved at low $\nu$  but changes substantially at $\nu \simeq \pi/4$, where the second and third  levels experience an avoided crossing. At $\nu \simeq \pi/2$ the mixed tensor interaction is dominated by the 10-th rank contribution, and  the upper two A  levels become nearly degenerate. At larger $\nu$ the 6-th rank interaction starts to contribute again, and the levels separate, eventually  forming an inverted $\nu=0$ spectrum at $\nu = \pi$ as expected.

\paragraph*{Mixing angle fitting}\label{SI:mixing_angle_fit}

To fit the mixing angle $\nu$ we perform a point cloud registration between measured defect plot and calculated $H^{(6+10)}_\text{tensor}(\nu)$ eigenvalue spectra. Unlike standard least-squares fitting, this algorithm is robust to missing or added peaks. The following procedure should also be robust to gradual scaling of the ro-vibrational fine structure as a function of $J$. We repeat the following for sections i)-iii) of the defect plot. 
\begin{enumerate}
\item Obtain the mean defect, referred to as $a(J)$ in main text, from the measured defect plot. Apply a $\Delta J = 7$-point moving average  (Figure~\ref{fig:SI_optimize_nu_pcreg}A).
\item Scale the calculated tensor eigenvalue spectrum by 0.007~cm$^{-1}$ and offset it by adding $a(J)$ from the previous step to make it resemble the measured defect plot as closely as possible; this step is heuristic.
\item Run the point cloud registration algorithm between the measured defect point cloud and processed calculated spectrum point cloud. Point cloud registration only shifts the energy of a point, without changing its $J$ value (in other words, the ``displacement field" vectors point along the vertical eigenvalue axis). This is valid because of the good agreement between measured peak counts and calculated nuclear spin weights. The displacement field vectors indicate how far and in which direction each measured defect must move to match the processed calculated spectrum. 
\item Generate a two-dimensional surface from the signed component of the displacement field vectors along the eigenvalue axis. Fit this surface to a polynomial that is second order in both $J$ and in the eigenvalue direction. 
\item Obtain the rms error of this fit. The best-fit mixing angle is the one that produces minimal error. Optimization curves for the three sections are shown in Figure~\ref{fig:SI_optimize_nu_pcreg}B.
\end{enumerate}
Note that by writing Equation~\ref{eq:fitting_fn} we have assumed a $J$-independent mixing angle. Formally, this requires that $\langle J ||T^{(6)}|| J\rangle$ and $\langle J ||T^{(10)}|| J\rangle$ scale identically with $J$. However, our data cannot conclusively establish relative $J$-scalings due to the relatively short continuous $J$-intervals and closeness of the mixing angle to ``pure" values $\nu = n\pi/2$.   %In reality, this is unlikely to be the case. However, even if they were to differ by $\sim J^3$, 
%for rather large differences in $\langle J ||T^{[6]}|| J\rangle$ and $\langle J ||T^{[10]}|| J\rangle$ J-scalings, 
%we would not necessarily be able to detect changes in the mixing angle 
%noise in the measured defect plot (e.g., missing or added peaks, and J-dependent scalings and offsets) 
There is also evidence of mixing in both sections ii) and iii) near $J=160$, which is not accounted for in the parametrization. %The pure rotational tensors we have used in this work may be slightly distorted compared to the ``actual" splitting pattern in our measurement, which must arise from ro-vibrational tensors . 

\paragraph*{b(J) fitting}

First, we fix the mixing angle $\nu$ to the optimal value determined in the previous section. Fitting $b(J)$ uses a standard least squares minimization, but is additionally robust to missing or added defect points. We again use the point cloud registration algorithm between the theoretical and measured defect spectrum to automatically identify corresponding points between the two spectra. The algorithm returns a displacement field and the overall rms error between the registered theoretical eigenvalue spectra and the measured defect. The squared ``Euclidean distance" between the theoretical eigenvalue spectra and the measured defects can therefore be obtained as 
\begin{equation}\label{eq:b(J)_sse}
    \textrm{sse} = \sum_{i=1}^N(\Delta y_i)^2 + N \sigma_\text{rms}^2
\end{equation}

where $N$ is the number of defect points involved in the fit, $\sigma_\text{rms}$ is the overall rms error after registration, and $\Delta y_i$ is the vertical component (along ``eigenvalue" axis) of the $i$th displacement vector in the displacement field. Equation~\ref{eq:b(J)_sse} is the expression that we would like to minimize.

To find the form of $b(J)$ we separately assume a constant, linear, and quadratic dependence on $J$. The rms fitting error for each of these three cases is given in Table~\ref{tab:fit_b(J)} for fitting of sections i)-iv). The data are consistent with a \textit{constant} $b(J)$ for each section, except for $J>247$, where the defect spectrum is no longer resolved. For each $J$ in this region, defects were obtained by fitting the raw spectrum to a cluster of Voigt profiles with constant amplitude and width. The number of Voigt profiles in each $J$-multiplet was determined from the calculated nuclear spin weight. We find that $b(J)$ exhibits quadratic, not constant, $J$ dependence in this region.

\paragraph*{Dark state landscape}
Two avoided crossings were previously identified in the R-branch spectrum at $J=215$ and 267~\cite{Changala2019}. We re-measure them at high SNR with the QCL and plot the results in Figure~\ref{fig:SI_Rbranch}. Panels B) and C) exhibit coupling strengths (bandwidth of the avoided crossings) of approximately $4\times 10^{-4}$~cm$^{-1}$. Combined with the avoided crossings observed in the P-branch in the main text, we obtain an average coupling strength of $\beta_\textrm{avg}\approx2\times10^{-2}~$cm$^{-1}$. Additionally, in a 4~cm$^{-1}$ frequency window we observe six avoided crossings. Combined with the triply degenerate $T_{1u}$(3) manifold, this implies an effective density of states $\rho_\textrm{obs}\approx2/$cm$^{-1}$.

\clearpage
\newcommand{\beginsupplementTabs}{%
        \setcounter{table}{0}
        \renewcommand{\thetable}{S\arabic{table}}%
     }
\beginsupplementTabs

\begin{table}
\caption{\textbf{Fitting $b(J)$ for each section for different $J$-dependencies.} Best-fit values for the scalar, linear, and quadratic coefficients are listed for each section. The rms error from the fitting is listed at the bottom of each section row. Section labelling follows those in Figure~\ref{fig:defect_plot}B. All units are in cm$^{-1}$. A \textit{constant} $b(J)$ provides nearly the best fit for all sections except $J>247$, which exhibits quadratic $J$-dependence. We found that adding a cubic fitting term $\propto J^3$ didn't reduce the rms error.}
\centering
\label{tab:fit_b(J)}
\begin{tabular}{ |p{2cm}||p{3cm}|p{3cm}|p{3cm}|  }
 \hline
 \multicolumn{4}{|c|}{$b(J)/\textrm{cm}^{-1}=$} \\
 \hline
 Section& $b_0\times 10^{-3}$ &$b_0\times 10^{-3}+$&$b_0\times 10^{-3}+$\\ & &$b_1\times 10^{-5}J$ &$b_1\times 10^{-5}J+$\\
 & & &$b_2\times 10^{-8}J^2$\\
 \hline
i)   &6.41    &7.77&7.93\\
& &-1.40 &-1.72 \\
& & &1.50\\
\hline
\textbf{$\sigma_\text{rms}\times 10^4$}&\textbf{2.14}  &\textbf{2.20}   &\textbf{2.20}\\
\hline
ii)   &7.13    &7.46&7.73\\
&    &-1.22& -0.341\\
&    & &0.382\\

\hline
\textbf{$\sigma_\text{rms}\times 10^4$}&\textbf{6.46}  &\textbf{6.92}   &\textbf{6.52}\\
\hline
iii)   &7.15    &8.61&9.42\\
&    &-0.755& -1.17\\
&    && -0.212\\
\hline
\textbf{$\sigma_\text{rms}\times 10^4$}&\textbf{2.58}  &\textbf{2.73}   &\textbf{2.83}\\
 \hline
 iv)   &10.0    &10.0&10.0\\
&    &-0.0025&-0.0031\\
&    &&-0.0007\\
\hline
\textbf{$\sigma_\text{rms}\times 10^4$}&\textbf{6.96}  &\textbf{6.96}   &\textbf{6.96}\\
 \hline
  $J>247$   &4.3    &8.6 &8.3\\
&    &-1.6& 4.9\\
&    &&-24\\
\hline
\textbf{$\sigma_\text{rms}\times 10^4$}&\textbf{2.78}  &\textbf{2.48}   &\textbf{1.53}\\
 \hline
\end{tabular}
\end{table} 

\clearpage

%Because of the high degeneracy of irreps in I$_h$, the symmetry-selected density of states for a single parity is always about 1/2 the total density of states in both cases.
\newcommand{\beginsupplementFigs}{%
        \setcounter{figure}{0}
        \renewcommand{\thefigure}{S\arabic{figure}}%
     }
     
\beginsupplementFigs

 \begin{figure}[t]
\begin{center}
\includegraphics[height=0.5\textheight]{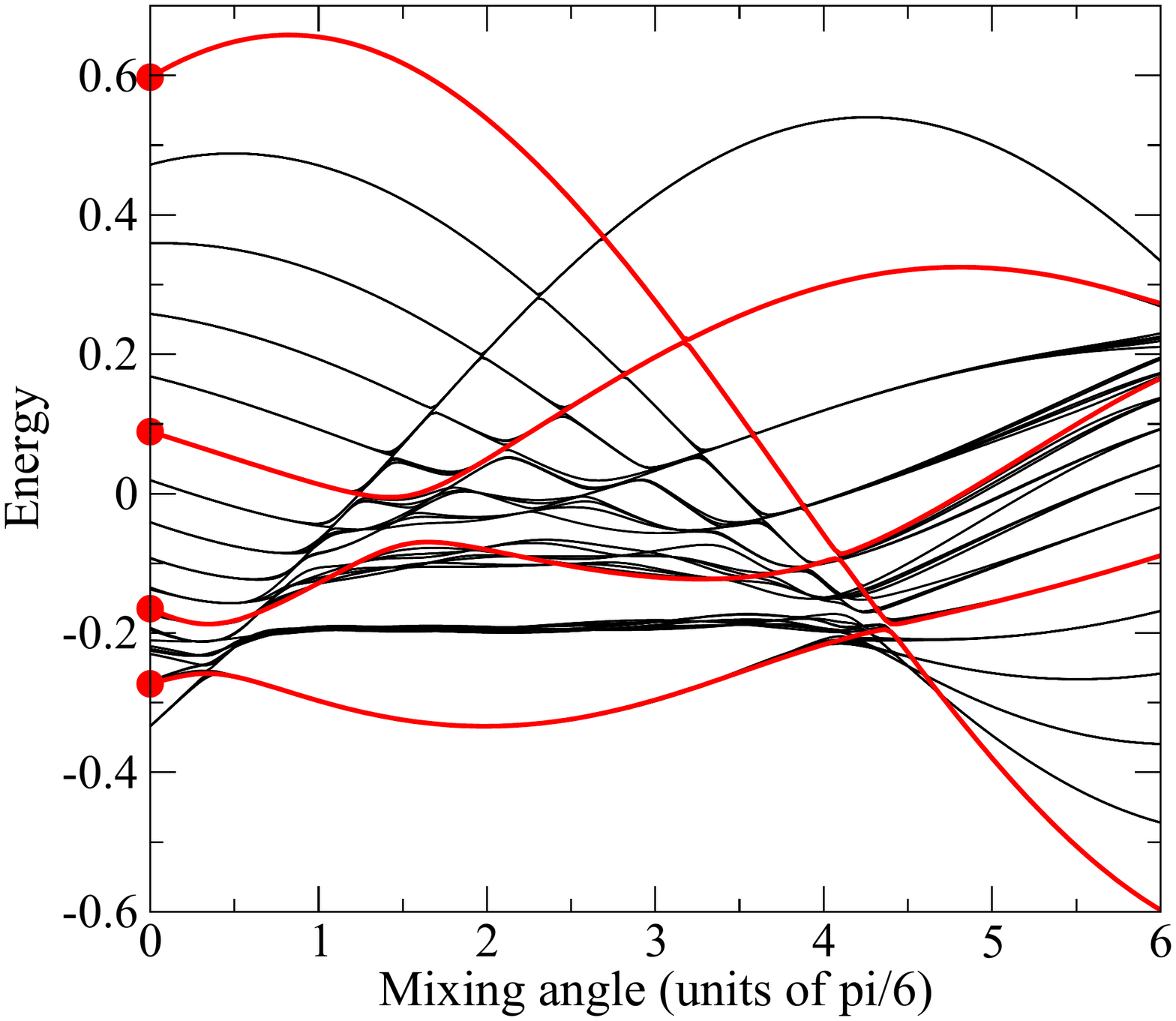}
\end{center}
\caption{Eigenvalues of the mixed-rank tensor Hamiltonian for $J=100$.  The A symmetry levels obtained by diagonalizing $\mathbf{H}_A$ in Eq.~(\ref{H_A}) are shown as red lines. The 6-th rank  A levels selected at $\nu=0$ are marked by red dots.}
% The spin-nonconserving reaction is neglected in this work.
\label{fig:SI_A_selection}
\end{figure}

\begin{figure}
\centering
	\includegraphics[width=\textwidth,scale=1]
{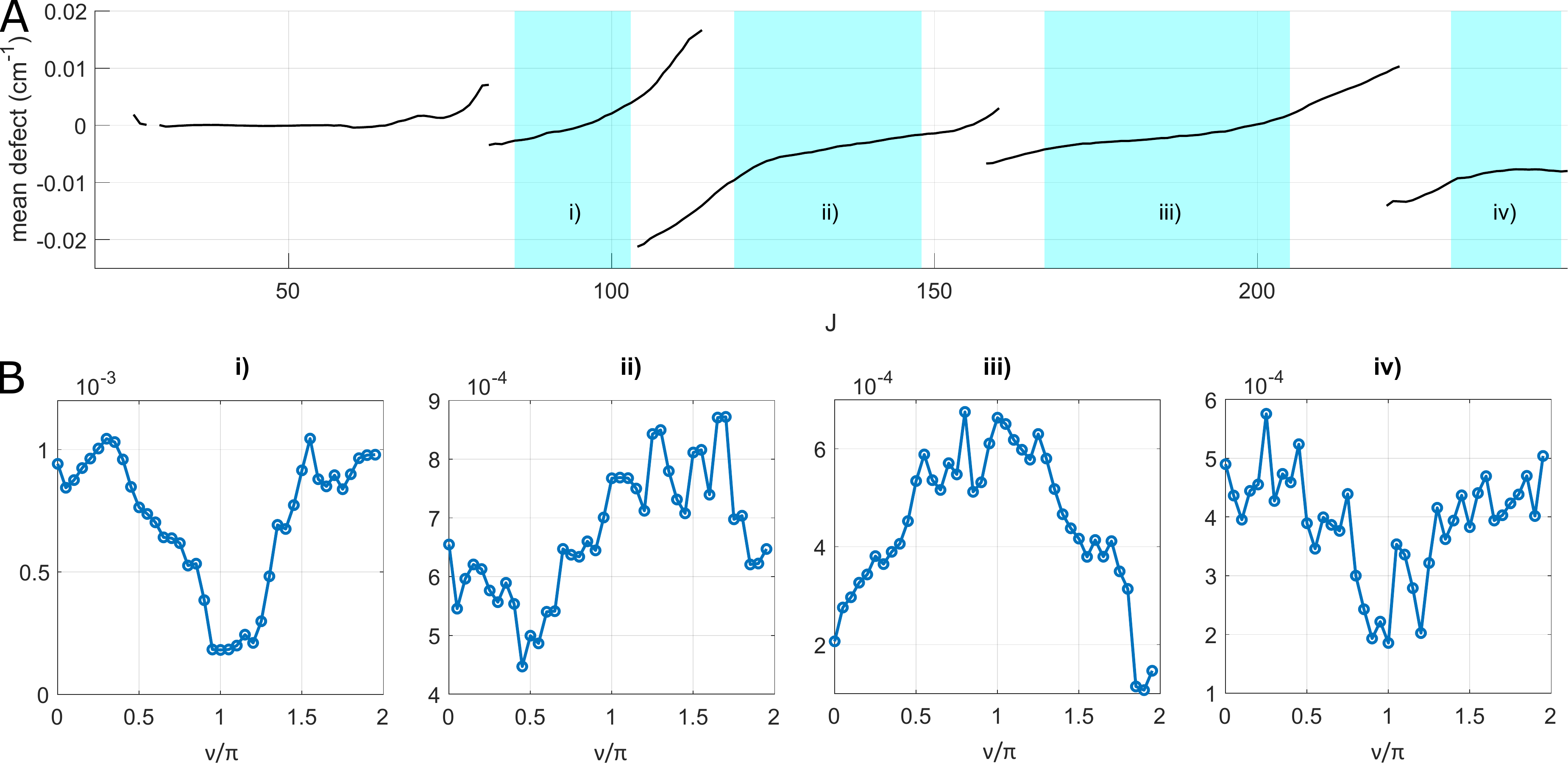}
	\caption{\textbf{A) $a(J)$ for the five sections in the defect plot.} $a(J)$ for each of the five sections is derived from their $J$-dependent mean defect, with a 7-point moving average. \textbf{B) rms error from fitting the mixing angle to sections i)-iv).} The best-fit $\nu$ is obtained by finding the minimum rms error using point cloud registration-based technique (see text). The optimal values are $\nu/\pi=$ 1.0, 0.45, 1.9, and 1.0 for sections i), ii), iii), and iv), respectively.}
	\label{fig:SI_optimize_nu_pcreg}
\end{figure}

\begin{figure}
	\includegraphics[trim={0cm 0 0cm 0},clip,width=\columnwidth,scale=1]{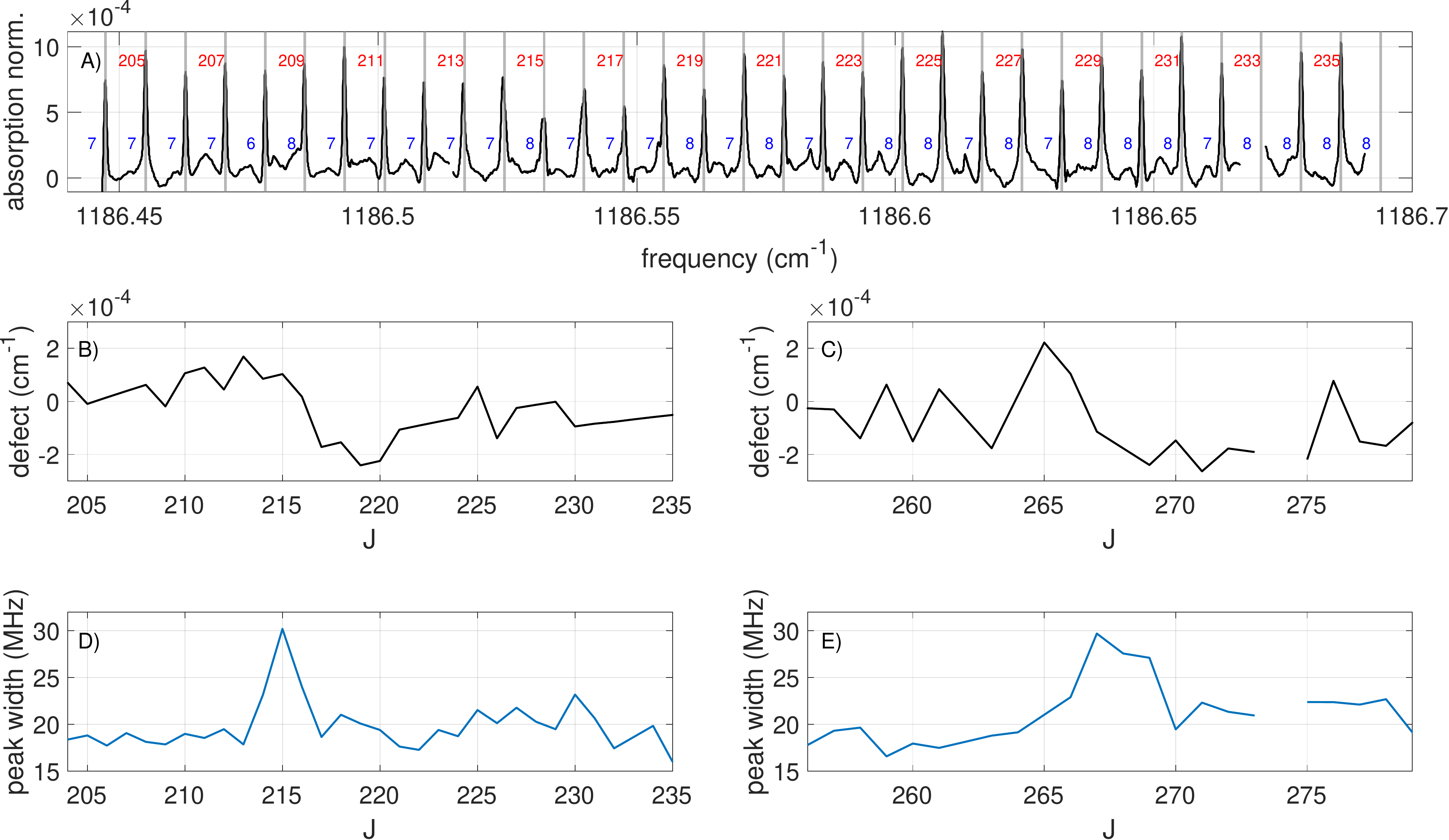}
	\caption{\textbf{Avoided crossings in the $T_{1u}$(3) R-branch.} A) Raw absorption spectrum around the avoided crossing at $J=215$. Red numbers indicate $J$ assignment, while blue numbers indicate calculated nuclear spin weight. B, C) Defect plots around avoided crossings at $J=215$ and 267, respectively~\cite{Changala2019}, showing the strength of the vibrational coupling to be approximately $4\times10^{-4}~$cm$^{-1}$. %The dispersion patterns in the plots indicate coupling strengths of $12$ and $14$MHz to the vibrational dark states, respectively. 
    D, E) Peak widths at $J=215$ and 267, respectively.}
	\label{fig:SI_Rbranch}
\end{figure} 
 
\end{document}